\documentclass[]{article}
\usepackage{titlesec} 

\renewcommand{\appendix}{
    \setcounter{section}{0}
    \renewcommand{\thesection}{Appendix \Alph{section}}
    \renewcommand{\thesubsection}{\thesection.\arabic{subsection}}
}
\usepackage{amsfonts,amssymb,amsmath,array}
\usepackage{graphicx}
\usepackage{amsbsy}
\usepackage{color}
\usepackage{hyperref} 
\usepackage{enumerate}
\usepackage{mathtools}
\usepackage{lipsum} 
\usepackage{cleveref} 
\usepackage{changepage} 
\usepackage[a4paper, total={7in, 10in}]{geometry}
 \usepackage{tabularx}
\usepackage{subfigure}
\usepackage{physics}
\usepackage{bm}
\usepackage[normalem]{ulem}
\usepackage{verbatim}
\usepackage{nccmath}
\usepackage{algorithm2e}
\RestyleAlgo{ruled}
\usepackage{nicematrix}
\usepackage{siunitx}
\usepackage{booktabs}

\def\ie{{i.e. }}

\def\de{\mathrm d}

\def\to{\rightarrow}

\newcommand{\vectheta}{\kappa}

\newcommand{\moy}[1]{\left\langle  #1 \right\rangle }

\begin{document}

\title{Optimal Control of an Electromechanical Energy Harvester}

\author{Dario Lucente$^1$, Alessandro Manacorda$^{2,*}$, Andrea Plati$^3$, Alessandro Sarracino$^4$ and Marco Baldovin$^2$}
\date{%
    {\small $^1$ Department of Mathematics \& Physics, University of Campania “Luigi Vanvitelli”, Viale Lincoln 5, 81100 Caserta, Italy\\%
    $^2$ CNR, Institute for Complex Systems, Universit\`a Sapienza, P.le A. Moro 5, 00185, Rome, Italy\\
    $^3$ Université Paris-Saclay CNRS Laboratoire de Physique des Solides, 1 rue Nicolas Appert, 91405, Orsay, France\\
    $^4$ Department of Engineering, University of Campania “Luigi Vanvitelli”, Via Roma 29, 81031 Aversa, Italy\\
    $^*$ Correspondence: alessandro.manacorda@isc.cnr.it}  
}

\maketitle

\begin{abstract}
Many techniques originally developed in the context of deterministic control theory have been recently applied to the quest for optimal protocols in stochastic processes. Given a system subject to environmental fluctuations, one may ask what is the best way to change in time its controllable parameters in order to maximize, on average, a certain reward function, while steering the system between two pre-assigned states.
In this work we study the problem of optimal control for a wide class of stochastic systems, inspired by a model of energy harvester. The stochastic noise in this system is due to the mechanical vibrations, while the reward function is the average power extracted from them. We consider the case in which the electrical resistance of the harvester can be changed in time, and we exploit the tools of control theory to work out optimal solutions in a perturbative regime, close to the stationary state. Our results show that it is possible to design protocols that perform better than any possible solution with constant resistance.
\end{abstract}










\section{Introduction}

Control theory deals with systems whose evolution can be externally steered by varying the value of some parameters. By suitably choosing time-dependent protocols for them, it is possible to drive the system between pre-assigned end states: a typical goal is to do so while minimizing a certain cost function (or maximizing some sort of reward). The theory was initially developed for deterministic systems~\cite{bellman1959mathematical,pontryagin1987mathematical,kirk2004optimal}, and later extended to quantum dynamics involving probabilistic descriptions: in this case even reaching the pre-assigned final state, which is given in terms of a probability distribution, is a non-trivial problem (usually referred to as \textit{shortcut to adiabaticity})~\cite{guery2019}. During the last two decades these ideas have found wide application in the context of stochastic systems~\cite{guery2023driving, blaber2023optimal}. The paradigmatic case of colloidal particles trapped in confining potentials has been studied both in the overdamped~\cite{schmiedl2007,MaPeGuTrCi2016, patron2022thermal} and underdamped regime~\cite{gomez2008optimal, muratore2014extremals, muratore2014nanomechanical, baldovin2024optimal}, also in the presence of non-harmonic confinements~\cite{plata2021, sanders2024optimal, sanders2024minimal}. Recent studies have extended the focus to out-of-equilibrium systems, including the so-called Brownian Gyrator~\cite{baldassarri2020}, stochastic dynamics with non-Gaussian~\cite{baldovin2022} and non-Markovian noise~\cite{loos2024universal}, stochastic resetting~\cite{de2023resetting,goerlich2024resetting}, active particles~\cite{shankar2022optimal, baldovin2023, davis2023active, garcia2024optimal}, granular materials~\cite{prados2021, ruiz2022optimal} and friction~\cite{plati2024control}. All these results strongly suggest that the framework of control theory is versatile enough to address complex real-life applications, such as those related to the optimization of energy harvesting in the presence of random fluctuations.

Energy harvesters are devices that can collect available energy from environmental sources, exploiting a huge gamma of different physical mechanisms~\cite{singh2021energy}.
The optimization of energy harvesting strategies represents a central issue for the maximization of the extracted power. 
From the theoretical perspective, for the mathematical description of these systems a central role is played by stochastic models, which mimic the random nature of fluctuations of the often unpredictable and uncontrolled environmental conditions. 

In this work we consider a linear stochastic model based on an underdamped Langevin equation that has been shown to describe with very good accuracy two kinds of energy harvesters, based on piezoelectric~\cite{halvorsen2008energy,costanzo2021stochastic,costanzo2023inference,sanfelice2024stochastic} and electromagnetic~\cite{Costanzo22entropy} effects. Piezoelectric materials~\cite{clementi2022review} can convert mechanical stress into electrical energy, and therefore they are used to design devices to harvest energy from environment vibrations. For instance the small currents produced can be exploited to feed sensors in wireless sensor networks. Electromechanical energy harvesters~\cite{shen2019unify} are typically realized with a permanent magnet, connected to the housing by a spring, and subjected to vibrations. The relative displacement with respect to a coil fixed to the housing and connected to an electrical load produces the conversion of the mechanical energy into electrical energy. Electromechanical harvesters are the electrical counterpart of ratchets, \ie devices able to convert environmental fluctuations into directed motion~\cite{Astumian94prl,Gnoli13prl,Manacorda14cmp}; the optimization problem equally applies for ratchets~\cite{Tarlie98pnas,Zulkowski15pre}, and the strategy introduced in this work is a suitable candidate for this challenge.

At variance with previous works~\cite{halvorsen2008energy,costanzo2021stochastic,Costanzo22entropy}, where optimization at fixed parameters was studied, we assume here that the resistance appearing in the load circuit of the energy harvester can be varied in time (i.e., we imagine to substitute it with a controllable potentiometer). We derive quite general equations for the optimal control of a class of stochastic processes that includes the considered model as a particular case. We then specialize to the case of small changes in the control parameter: this choice allows us to solve these equations perturbatively. Remarkably, we are able to identify new control protocols that perform better than the previously known optimal strategy, obtained by suitably fixing a constant value of the resistance. Our present results are obtained in a particular case, but they serve as a proof of principle for future, more exhaustive investigations.

The structure of the paper is as follows. In Sec.~\ref{sec:harvester} we discuss the stochastic model describing the energy harvester that we want to optimize.
In Sec.~\ref{sec:pontryagin} we introduce the concepts of control theory that we need in order to maximixe the performance, and we carry on explicit calculations for a wide class of stochastic systems, including the one introduced before. A perturbative method for the solution of the problem is outlined.
Section~\ref{sec:results}  is devoted to the discussion of the results we get for some particular choices of the parameters. Finally in Sec.~\ref{sec:discussion} we draw our conclusions.
\section{Energy harvester dynamics}
\label{sec:harvester}

We consider a theoretical model that has been shown to very well describe the behavior of electromagnetic and piezoelectric energy harvesters driven by broadband vibrations, not only for the average values of the main quantities of interest, but even at the fine level of their fluctuations~\cite{costanzo2021stochastic,Costanzo22entropy}. 
In the simplest realization, both these systems consist of a movable part (permanent magnet or piece of piezoelectric material, respectively) that is subjected to vibration and is electromechanically coupled with the current flowing in the output electric circuit. We will focus here on the electromagnetic harvester, but our study can be easily extended to the piezoelectric case. In particular, the main variables appearing in the theoretical model are the magnet position $x$ with respect to the coil, its velocity $v$ and the current $I$. 
The forces acting on the magnet are due to a harmonic potential (representing the spring), the viscous friction of the air, the coupling with the current (Lorentz force, due to the interaction between the current $I$ flowing into the coil and the induction field of the magnet) and the stochastic driving, which we describe as a white noise.
Therefore, in the linear approximation, the system is described by an underdamped Langevin equation for the magnet position, coupled with a deterministic equation which describes the time evolution of the current:
\begin{equation}\label{eq:harvester}
    \begin{aligned}
        \dot x &= v \\
        M \dot v &= -k_s x - \gamma v - \theta I + M \xi \\
        L_C \dot I &= \theta v - (R_C+R) I \\
        \moy{\xi(t)\xi(t')} &= 2 D_0 \delta(t-t').
    \end{aligned}
\end{equation}
In the above equations $M$ represents the mass of the magnet, $\gamma$
 represents the viscous damping due to the air friction, $k_s$
 is the elastic constant of the spring system, $I$ is the current at the electrical terminals, $L_C$ and $R_C$
 are the coil inductance and resistance, respectively, $R$ is the load resistance of the output circuit and $\theta$
 is the effective electromechanical coupling factor, taking into account coil geometrical properties, number of turns and magnetic field strength. The coefficient $D_0$ quantifies the noise amplitude.
Finally, we note that the coupling between magnet velocity and electrical current can be interpreted as a feedback mechanism, and the model can be mapped into a single generalized Langevin equation with memory effect~\cite{costanzo2021stochastic}.


The extracted power is obtained from the average of the square of the current flowing in the load resistance. Here we consider the case of a time-dependent value of $R=R(t)$ 
\begin{equation}\label{eq:power}
    P_{harv}[R(t)] = \int^{tf}_{t_0} \de t \, R(t) \langle I^2(t)\rangle. 
\end{equation}
Our goal is to find the optimal protocol $R(t)$ which maximizes the extracted power.

\section{Global optimization}
\label{sec:pontryagin}

A global optimal control of the problem introduced above can be found through the Pontryagin's \
Maximum Principle (PMP)~\cite{pontryagin1987mathematical, kirk2004optimal, Liberzon}. We will see that this strategy leads to a quite involved system of ordinary differential equations, which can be studied perturbatively by taking the stationary optimal solution as a 
reference point and expanding the dynamics around the corresponding stationary state.

\subsection{Pontryagin's Maximum Principle}

Let's consider a dynamical system characterized by its \textit{state} $x \in \mathbb{R}^d$, following the evolution equation
\begin{equation}\label{eq:dynamics}
\dot x (t) = F (x (t), u (t)) \ , \quad x (t_0) = x_0 \ , \quad x(t_f) = x_f \ ,
\end{equation}
being $u \in U \subset \mathbb{R}$ the \textit{control} applied to the system. The initial and final conditions, $x_0$ and $x_f$ respectively, are fixed, as well as the final time $t_f$. These constraints identify a fixed endtime, fixed endpoint problem, and the control $u(t)$ must be able to satisfy them. One then aims at maximizing a \textit{reward} functional $J$ defined as
\begin{equation}\label{eq:reward1}
    J[u] = -\int_{t_0}^{t_f} \de t \, L(x(t),u(t)) \ .
\end{equation}
The reward is a functional of the applied control $u(t)$; the \textit{Lagrangian} term $L(x,u)$ can depend on the control explicitly and implicitly through $x(t)$.

The PMP provides the necessary conditions for the global optimality of the control $u(t)$: given the dynamics and the reward defined above, one can typically define an Hamiltonian function $H(x,\lambda,u)$ as
\begin{equation}\label{eq:hamiltonian}
    H(x,\lambda,u) = \lambda \cdot F(x,u) - L(x,u) \ , 
\end{equation}
being $\lambda \in \mathbb{R}^d$ called the \textit{costate} variable of the problem. The PMP states that if $u^*(t)$ is a globally optimal control, the following equations hold:
\begin{equation}\label{eq:PMP}
    \begin{aligned}
        \dot x^*(t) &= + \partial_\lambda H(x^*(t),\lambda^*(t),u^*(t)) = F(x^*(t),u^*(t)) \ , \quad &\text{(dynamics)} \\
        \dot \lambda^*(t) &= - \partial_x H(x^*(t),\lambda^*(t),u^*(t)) \ , \quad &\text{(costate)} \\
        H(x^*(t),\lambda^*(t),u^*(t)) &\geq H(x^*(t),\lambda^*(t),u) \quad \forall u \in U \ . \quad &\text{(control)}
    \end{aligned}
\end{equation}
Here $x^*(t)$ and $\lambda^*(t)$ correspond to the optimal state and costate evolved under the action of the optimal control $u^*(t)$. The last equation expresses the \textit{local} optimality condition for $u$; if the maximum does not lies on the boundary $\partial U$, it can be rewritten as $\partial_u H(x^*(t),\lambda^*(t),u^*(t))=0$.

\subsection{PMP for affine dynamics}

We are now ready to find the optimal control $u(t)$ for the electromechanical harvester introduced in Sec.~\ref{sec:harvester}. Before doing so, we move from the space of stochastic variables to the deterministic evolution of equal-time correlations. The dynamics in Eq.~\eqref{eq:harvester} can be written in the form
\begin{equation}\label{eq:OU}
    \dot X(t) = - A(u(t)) X(t) + B \xi(t) \ , \quad \moy{\xi_\mu(t) \xi_\nu(t')} = \delta_{\mu\nu} \delta(t-t'),
\end{equation}
being $X \in \mathbb{R}^n$ and $A, B \in \mathbb{R}^{n\times n}$. We here study the case where $B$ is constant and $A(u)$ is a first degree polynomial in $u$, namely 
\begin{equation}
\label{eq:polyA}
A(u) = A_0 + u A_1 \,.
\end{equation}
The evolution is linear: this means that if the initial state is a Gaussian distribution centered in zero, for any later time this property will be preserved. As a consequence, the state of the system is completely determined by the covariance matrix $\Sigma_{\mu\nu}(t) = \moy{X_\mu(t)X_\nu(t)}$, whose evolution reads~\cite{Gardiner}
\begin{equation}\label{eq:cov}
    \dot \Sigma(t) = -\left[ A(u(t)) \Sigma(t) + \Sigma(t) A^T(u(t)) \right] + 2 D \ , \quad 2D = B B^T \ .
\end{equation}
From now on, we will focus on the covariance vector $\sigma \in \mathbb{R}^d$ consisting of the $d=n(n+1)/2$ independent components of the matrix $\Sigma$. Eq.~\eqref{eq:cov} leads to
\begin{equation}\label{eq:sigma-dyn}
    \dot \sigma(t) = - M(u(t)) \sigma(t) + b \ ,
\end{equation}
where the matrix $M(u)$ and the vector $b$ are univocally determined by $A$ and $D$, respectively. Eq.~\eqref{eq:sigma-dyn} defines the dynamics of our system, determined by its state $\sigma$.
The extracted power defined in Eq.~\eqref{eq:power} represents the reward of our problem. Since it depends linearly on the equal-time correlations, it can be written as
\begin{equation}\label{eq:reward}
    J[u] = \int^{t_f}_{t_0} \de t \, u(t) \, \vectheta \cdot \sigma(t) \ ,
\end{equation}
being $\vectheta$ a constant vector. The Hamiltonian of the optimal problem then reads
\begin{equation}\label{eq:Hamiltonian}
    H(x,\lambda,u) = \lambda \cdot [- M(u) \sigma + b] + u \, \vectheta\cdot\sigma \ , \quad M(u) = M_0 + u M_1 \ ,
\end{equation}
being $M_0$ and $M_1$ constant matrices, because of~\eqref{eq:polyA}. The optimal problem is then determined by $M_0$, $M_1$, $b$ and $\vectheta$. Explicit expressions of these objects for the harvester model~\eqref{eq:harvester} are provided in~\ref{app:dimensions}; for the moment we keep the discussion at an higher level of generality.
The PMP equations read (omitting the asterisks and time dependence from now on)
\begin{subequations}
\label{eq:PMP-affine}
    \begin{align}
        \dot\sigma &= \partial_\lambda H = - M(u) \sigma + b \ , \label{eq:PMP-sigma} \\
        \dot\lambda &= - \partial_\sigma H = M^T(u) \lambda - u \vectheta \ , \label{eq:PMP-lambda} \\
        0 &= \partial_u H = - \lambda^T M_1 \sigma + \vectheta\cdot\sigma \ , \label{eq:PMP-u} \\
        \sigma(t_f) &= \sigma(t_0) = \sigma_s \ . \label{eq:PMP-BC}
    \end{align}
\end{subequations}
Here we included the boundary conditions~\eqref{eq:PMP-BC} for the covariance $\sigma$, defined by the same value $\sigma_s$. This value ideally represents a steady-state of the dynamics for $t<t_0$ and $t>t_f$, corresponding to the control value $u=u_s$. The control $u(t)$ is changed in the time interval $(t_0,t_f)$ and, with these prescriptions, we are guaranteed to start and to end in a stationary state purely determined by $u_s$, without any need of thermal relaxation. This is the main advantage of working with correlation functions, thanks to the linearity of the dynamics - \textit{i.e.} the Gaussianity of the state. The control is kept at $u\equiv u_s$ outside the considered time interval.

Equations~\eqref{eq:PMP-affine} constitute a system of $2d$ ordinary differential equations for the $2d+1$ variables $(\sigma,\lambda,u)$, with $2d$ boundary conditions for $\sigma$~\eqref{eq:PMP-BC}
and one constraint~\eqref{eq:PMP-u} given by the stationarity along $u$.
A typical strategy of solution is to use the constraint~\eqref{eq:PMP-u} to remove $u$ from the dynamics and obtain a closed system of $2d$ equations for $(\sigma,\lambda)$ from Eqs.~\eqref{eq:PMP-sigma} and~\eqref{eq:PMP-lambda}. Unfortunately the constraint does not depend on $u$, so it cannot be used directly to this aim. But since it must vanish for all $t \in (t_0,t_f)$, its time derivatives will also vanish and one has
\begin{equation}\label{eq:dt-constraint}
    0 = \frac{\de}{\de t}\partial_u H = \lambda^T \left[M_1,M_0\right] \sigma - \vectheta^T M_0 \sigma - \left( \lambda^T M_1-\vectheta^T \right) b \ ,
\end{equation}
and
\begin{equation}\label{eq:d2t-constraint}
    \begin{aligned}
        0 &= \frac{\de^2}{\de t^2} \partial_u H = \lambda^T \left[ M , \left[ M_1,M_0 \right] \right] \sigma + \lambda^T\left( \left[ M_1,M_0 \right] - M M_1 \right) b - \vectheta^T \left( u \left[ M_1,M_0 \right] - M_0 M \right) \sigma - \vectheta^T (M_0-uM_1) b \\
        &= \lambda^T \left[ M_0,\left[ M_1,M_0\right]\right] \sigma - \lambda^T(2M_0M_1 - M_1 M_0 )b + \vectheta^T M_0^2 \sigma - \vectheta^T M_0 b + \\
        &\quad + u \left[ \lambda^T \left[M_1,\left[M_1,M_0\right]\right] \sigma - \lambda^T M_1^2 b + \vectheta^T (2M_0 M_1 - M_1M_0) \sigma + \vectheta^T M_1 b \right] \ .
    \end{aligned}
\end{equation}
Eq.~\eqref{eq:d2t-constraint} gives us a \textit{pointwise} relation for the optimal control $u(t) = u(\sigma(t),\lambda(t))$. 
The optimal solution $u$ is clearly nonlinear in $(\sigma,\lambda)$, meaning that also Eqs.~\eqref{eq:PMP-sigma}-\eqref{eq:PMP-lambda} will become nonlinear as soon as we plug the expression for the control in them: this is the price we pay for eliminating $u$ from the equations. We will see in Sec.~\ref{sec:perturbative} a perturbative approach to tackle this problem.


\subsection{Is the stationary optimum a global optimum?}
\label{sec:statdyn}

We know from Ref.~\cite{Costanzo22entropy} that, if one only considers a stationary control $u(t) \equiv u_s$, there exists a stationary optimum $u^*$ maximizing the stationary harvested power; the latter reads $j(u_s) = L(\sigma_s,u_s)/(t_f-t_0) = u_s \, \vectheta\cdot\sigma_s$, with $\sigma_s = M^{-1}(u_s)b $ in our notation. The optimal $u^*$ then satisfies
\begin{equation}\label{eq:ss-optimum}
    \left. \frac{\de j}{\de u_s} \right\vert_{u_s=u^*}= 0 \quad \Leftrightarrow \quad \left.  \vectheta^T \left( I - u_s M^{-1}(u_s) M_1 \right) \sigma_s \right\vert_{u_s=u^*} = 0 \ .
\end{equation}
This relation determines a constant optimal $u^*$ and we now wonder if this optimum 
can also be an optimum of the dynamic control problem, when $u(t)$ is allowed to 
change in time. We then look at the stationary dynamics with boundary conditions $\sigma(t_0) = \sigma(t_f) = \sigma_s \vert_{u_s=u^*}$. Substituting $u(t)\equiv u^*$ into PMP Eqs.~\eqref{eq:PMP-affine}, 
one has
\begin{equation}
    \sigma(t) \equiv \sigma^* = M ^{-1}(u^*) b \ , \quad \lambda^T(t) \equiv {\lambda^*}^T = u^* \vectheta^T M^{-1}(u^*) \ ,
\end{equation}
and the constraint then reads
\begin{equation}\label{eq:ss-constraint}
    0 = \partial_u H = - {\lambda^*}^T M_1 \sigma^* + \vectheta\cdot\sigma^* = \vectheta^T \left( I - u^* M^{-1}(u^*)M_1 \right) \sigma^* \ .
\end{equation}
Since Eq.~\eqref{eq:ss-constraint} is the same condition satisfied by $u^*$ in Eq.~\eqref{eq:ss-optimum}, the stationary optimum fulfills Pontryagin's conditions and is a candidate to be an optimal control for the dynamic problem. However, due to the non-uniqueness of the solutions, it may happen that other protocols perform even better. We will assess the optimality of the stationary protocol under specific conditions in Sec.~\ref{sec:results}.

\subsection{A perturbative approach to the solution}
\label{sec:perturbative}
As anticipated in Sec.~\ref{sec:statdyn}, the standard procedure for finding optimal protocols involves expressing $u$ as a function of $\sigma$ and $\lambda$ and subsequently solving the resulting nonlinear boundary value problem. Note that, when eliminating $u$ from the system by mean of Eq.~\eqref{eq:d2t-constraint}, we are not really enforcing the constraint~\eqref{eq:PMP-u}: the latter is fulfilled if Eq.~\eqref{eq:d2t-constraint} holds $\forall t \in (t_0,t_f)$ \textit{and} Eqs.~\eqref{eq:PMP-u},~\eqref{eq:dt-constraint} are verified for at least one value of $t$ in the same time interval. This means that we have two more constraints (local in time), and the system~\eqref{eq:PMP-affine} is overdetermined. This situation is quite well known in the context of optimal control for underdamped systems~\cite{gomez2008optimal, muratore2014extremals, baldovin2024optimal}, and can be solved formally by admitting impulsive variations of the control at the beginning and at the end of the protocol, in order to abruptly change the boundary conditions in such a way that they fulfill the constraints discussed above. While these dynamical discontinuities are physically unrealistic, several approaches to regularize them have been proposed~\cite{muratore2014extremals, muratore2014nanomechanical, muratore2017application, baldovin2023, baldovin2024optimal, plati2024control}. 
These regularizations allow to write down explicit protocols that are arbitrarily close to the optimal one and can be actually perfomed in experiments~\cite{baldovin2024optimal}. However, since the focus of this paper is mainly conceptual, we have chosen not to pursue these alternatives, for the sake of simplicity, leaving them for further studies.
We focus therefore on controls of the form
    \begin{equation}
        u(t) = u_0 \delta(t-t_0) + u_b(t) + u_f \delta(t-t_f) \ ,
    \end{equation}
being $u_b(t)$ the control in the bulk, $t\in(t_0,t_f)$, without the end times; $u_0 = -\log \Delta_0$ and $u_f = \log \Delta_f$ are instead the amplitudes of the control pulses at $t=t_0,t_f$ respectively, determining the discontinuities in $\sigma$ and $\lambda$. 

The task of solving the nonlinear problem~\eqref{eq:PMP-affine} may be very challenging in most situations.
A great simplification occurs if one considers the optimization problem for periodic conditions $\sigma(t_0) = \sigma(t_f) = \sigma_s = M^{-1}(u_s) b$ close to the optimal stationary solution $\sigma^* = M^{-1}(u^*)b$ discussed in Sec.~\ref{sec:statdyn}.
By eliminating $u(t)$ from Eqs.~\eqref{eq:PMP-sigma}$-$\eqref{eq:PMP-lambda} using Eq.~\eqref{eq:d2t-constraint}, and assuming that the optimal protocol remains close to $u^*$ at any time, we can expand the unknowns around the point $\sigma^* = M^{-1}(u^*)b$, $\lambda^* = u^* {M^T}^{-1}(u^*) \vectheta $ as
    \begin{equation}
        \begin{aligned}
            \sigma(t) &= \sigma^*+\delta\sigma(t) \ , \\
            \lambda(t) &= \lambda^*+\delta\lambda(t) \ . \\
        \end{aligned}
    \end{equation}
    The linearized system is given by
    \begin{equation}\label{eq:linearized}
        \frac{\de}{\de t} 
        \begin{pmatrix}
            \delta\sigma(t) \\
            \delta\lambda(t)
        \end{pmatrix}
        = - W
        \begin{pmatrix}
            \delta\sigma(t) \\
            \delta\lambda(t)
        \end{pmatrix}
        \ , 
        \end{equation}
   being $W\in \mathbb{R}^{2n \times 2n}$ determined by Eqs.~\eqref{eq:PMP-affine} and~\eqref{eq:d2t-constraint}. The solution of Eq.~\eqref{eq:linearized} for a generic initial condition $(\delta\sigma(t_0^+),\delta\lambda(t_0^+))$ reads
    \begin{equation}\label{eq:bulk-linear}
        \begin{aligned}
            \delta\sigma(t) &= U_{\sigma\sigma}(t,t_0) \delta\sigma(t_0^+) + U_{\sigma\lambda}(t,t_0) \delta\lambda(t_0^+) \ , \\
            \delta\lambda(t) &= U_{\lambda\sigma}(t,t_0) \delta\sigma(t_0^+) + U_{\lambda\lambda}(t,t_0) \delta\lambda(t_0^+) \ . \\
        \end{aligned}
    \end{equation}
    In the expression above, we have introduced the propagator of the dynamics 
    \begin{equation}\label{eq:UpropagatorU}
    U(t,t') = \exp{- W(t-t')} \ ,
    \end{equation}
    and its block representation
     \begin{equation}
        \quad U = 
        \begin{pNiceArray}{c|c}
            U_{\sigma\sigma} & U_{\sigma\lambda} \\
            \hline
            U_{\lambda\sigma} & U_{\lambda\lambda}
        \end{pNiceArray}
        \ .
    \end{equation}
 The boundary condition of the linearized dynamics are related to the imposed p.b.c. $\sigma_s$ via (see~\ref{app:discontinuity} for the derivation)
    \begin{equation}
        \delta\sigma(t_0^+) = e^{-M_1 u_0} \sigma_s - \sigma^* \ , \quad
        \delta\sigma(t_f^-) = e^{M_1 u_f} \sigma_s - \sigma^* \ .
        \label{eq:boundary-linear}
    \end{equation}

Now, Eq.~\eqref{eq:bulk-linear} together with Eq.~\eqref{eq:boundary-linear} define a \emph{linear} boundary value problem which can be formally solved.
As a result, we will obtain an initial value problem that satisfies the original boundary value problem without relying on numerical schemes such as the shooting method.

Using Eq.~\eqref{eq:bulk-linear} together with Eq.~\eqref{eq:boundary-linear}, we can determine $\delta\lambda(t_0^+)$ and $\delta\lambda(t_f^-)$ as a function of $u_0$ and $u_f$. The remaining step for solving the optimal control problem consists in enforcing the validity of the constraints $\partial_u H = \dfrac{\de}{\de t}\partial_u H = 0$ at $t=t_0^+$ and $t=t_f^-$. In the perturbative approach we consider here, these constraints can be reduced to a system of coupled algebraic equations. Indeed, by virtue of the Cayley-Hamilton theorem~\cite{Frobenius1878,lang1987linear}, every function $f(\cdot)$ of a $d\times d$ matrix $G$ can be expressed as a polynomial of degree $d-1$ in $G$, that is 
\begin{equation}
    f(G)=\sum_{l=0}^{d-1}g_l G^l\,.
\end{equation} 

Thus, by inserting $f(\cdot)=\exp(\cdot)$ and $G=u_0 M_1$ or $G=u_f M_1$ in Eq.~\eqref{eq:boundary-linear}, we get
\begin{align}
    &\delta\sigma(t_0^+)=\sum_{l=0}^{d-1}(-1)^l m_l M_1^l\sigma_s u_0^l -\sigma^*\,,\label{eq:poly-sigma_0}\\
    &\delta\sigma(t_f^-)=\sum_{l=0}^{d-1} m_l M_1^l\sigma_s u_f^l -\sigma^*\,,\label{eq:poly-sigma_f}
\end{align}
being $m_l$ the expansion coefficients of the exponential function. Solving Eq.~\eqref{eq:bulk-linear} to express $\delta\lambda(t_0^+)$ and $\delta\lambda(t_f^-)$ as functions of $\delta\sigma(t_0^+)$, $\delta\sigma(t_f^-)$, and substituting Eqs.~\eqref{eq:poly-sigma_0}-\eqref{eq:poly-sigma_f} for the latter, we obtain
\begin{align}
    &\lambda(t_0^+)=\sum_{l=0}^{d-1}\left[ \lambda_l^{(+,0)} u_0^l +  \lambda_l^{(+,f)} u_f^l\right]\,,\label{eq:poly-lambda_0}\\
    &\lambda(t_f^-)=\sum_{l=0}^{d-1}\left[ \lambda_l^{(-,0)} u_0^l +  \lambda_l^{(-,f)} u_f^l\right]\,, \label{eq:poly-lambda_f}
\end{align}
where $\lambda_l^{(\pm,0)}$ and $\lambda_l^{(\pm,f)}$ are vectorial coefficients. 
Since $\delta\sigma(t_0^+)$, $\delta\sigma(t_f^-)$, $\delta\lambda(t_0^+)$ and $\delta\lambda(t_f^-)$ are all polynomials of degree $d-1$ {\it at most} in $u_0$ and $u_f$, and the constraints $\partial_u H = \dfrac{\de}{\de t}\partial_u H = 0$ involve only quadratic nonlinearities, finding the optimal protocol reduces to solving two coupled algebraic equations of degree $2(d-1)$ at most, i.e. 

\begin{equation}
        \sum_{l=0}^{2(d-1)}c^{(1)}_l u_0^l u_f^{2(d-1)-l}= 0 \ , \quad
         \sum_{l=0}^{2(d-1)}c^{(2)}_l u_0^l u_f^{2(d-1)-l}= 0 \, ,
        \label{eq:system}
    \end{equation}
   being $c^{(1)}_l$ and $c^{(2)}_l$ coefficients determined by the problem (i.e. by the matrix $M$). 
In general the above system admits multiple solutions and the optimal control must be determined as the one that maximizes the reward
\begin{equation}
        J[u] = \int^{t_f}_0 \de t \, u(t) \vectheta\cdot\sigma(t) =
        \vectheta \cdot \left[ \left( e^{M_1 u_f} - e^{-M_1 u_0} \right) M_1^{-1} \sigma_s + \int^{t_f}_0 \de t \, u_b(t) \sigma(t) \right]\,,
        \label{eq:reward_linear}
    \end{equation}
where the first terms stem from discontinuities while the last from the bulk dynamics (see~\ref{app:discontinuity}). 
In the following, we will apply the aforementioned procedure to the energy harvester model~\eqref{eq:harvester} introduced in Sec.~\ref{sec:harvester}.

\section{Results}
\label{sec:results}
The framework introduced in Sec.~\ref{sec:pontryagin} to identify optimal protocols for affine dynamics can be applied to the energy harvester model~\eqref{eq:harvester}. A standard analysis of the physical dimensions involved (see~\ref{app:dimensions}) allows us to rewrite the evolution in dimensionless units as
\begin{equation}
\label{eq:nodimdyn}
    \begin{aligned}
        \dot x &= v \\
        \dot v &= -\alpha x - \beta v - I + \xi \\
        \dot I &= v - \varepsilon I \\
        \moy{\xi(t)\xi(t')} &= 2 \delta(t-t')\,,
    \end{aligned}
\end{equation}
with the control $u$ acting on
$$
\varepsilon=\zeta+u\,.
$$
Here $\alpha$, $\beta$ and $\zeta$ are constant parameters, while $u$ is a time-dependent control taking positive values. We assume that $u$ can be changed with arbitrary speed and precision: this is of course an idealization, since our ability to control $u$ actually depends on the operating range of the potentiometer in use. In~\ref{app:dimensions} we also show that the stationary extracted power in dimensionless units reads
\begin{equation}
    P_s(u_s) = \frac{u_s}{\varepsilon_s+\beta(1+\alpha+\varepsilon_s(\beta+\varepsilon_s))}
\end{equation}
(with $\varepsilon_s=\zeta+u_s$), and is maximized by
\begin{equation}
\label{eq:uopt-st}
u^* = \sqrt{\frac{\alpha  \beta +(\beta +\zeta)(1+\beta\zeta) }{\beta }}  \ .
\end{equation}
As discussed in Section~\ref{sec:statdyn}, the stationary protocol $u=u^*$ is a solution of the (dynamical) Pontryagin's optimal problem~\eqref{eq:PMP-affine}, with boundary conditions $\sigma(0)=\sigma^*$ and $\sigma(t_f)=\sigma^*$. In other words, if the system is initially prepared in the stationary state $\sigma^*$ and needs to be brought back to $\sigma^*$ in a time $t_f$, the intuitive strategy of keeping the potentiometer fixed at $u=u^*$ does fulfill Pontryagin's necessary condition (for any time interval $t_f$). However, the solution of problem~\eqref{eq:PMP-affine} is not unique in general, so there is no warranty that such prescription is actually the best. As a matter of fact, in the considered case better solutions can be found, at least for some choices of the parameters: we will come back to this point in the following Section~\ref{sec:alpha0}.

All considerations made in Section~\ref{sec:pontryagin} can be applied, in principle, to model~\eqref{eq:nodimdyn}. The explicit calculations may however turn out to be quite involved, due to the high dimensionality of system~\eqref{sec:statdyn}: here $\sigma$ and $\lambda$ are $6$-dimensional vectors, hence $W$ and $U$ turn out to be $12\times12$ matrices. Our goal here is not to find the solution of the optimal problem for a realistic choice of the parameters, but rather to provide a proof of principle of the usefulness of the method and the possibility to have dynamical protocols that are more efficient than the statical one $u=u^*$. Therefore, in the following we focus on a particular regime for which we are able to provide explicit solutions. A more systematic exploration of the several regimes of the model is left as the subject for future research work.

\begin{figure}
\includegraphics[width=.99\linewidth]{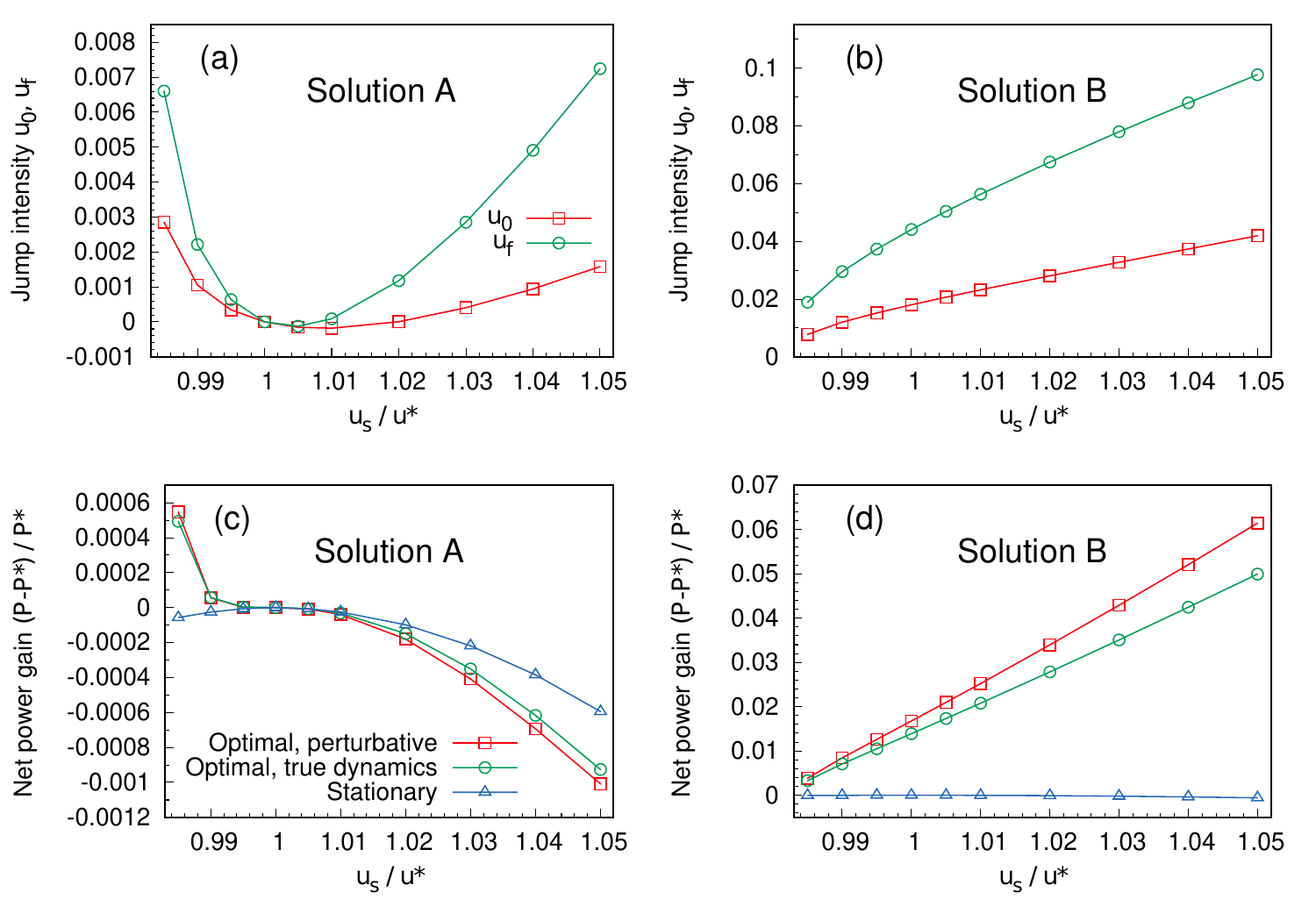}
\caption{Characterization of the solutions of PMP. Panels (a) and (b) show the intensity of the infinite discontinuities $u_0$ and $u_f$ occurring at the beginning and at the end of the protocol, in the two physically admissible solutions A and B of system~\eqref{eq:system}. Different choices of the stationary control $u_s$, fixing the boundary conditions~(\ref{eq:bc}), are considered. Panels (c) and (d) account for the corresponding net power gain (or loss), with respect to the stationary strategy $u=u^*$. The red squares refer to the value of the average power computed within the perturbative approach. Green circles are obtained by plugging the solution protocol $u$ into the original (non-perturbative) dynamics, and computing the average power of the process (see caption of Fig.~\ref{fig:dyn} for details). Of course in this case the final state $\sigma(t_f)$ will not match the prescribed boundary condition exactly - see Fig.~\ref{fig:dyn}. The distance between the two curves is an indicator of the quality of the perturbative approximation. Finally, the blue triangles represent, for reference, the power obtained with a stationary protocol $u=u_s$. Parameters: $\alpha=0$, $\beta=1$, $\zeta=2$, $t_f=0.25$.}
\label{fig:NetPower}
\end{figure}

\subsection{An explicit example: the case $\alpha=0$}
\label{sec:alpha0}
A considerable simplification in model~\eqref{eq:nodimdyn} arises when considering the particular case $\alpha=0$. From a physical point of view this choice corresponds to an overdamped limit, in which the relaxation dynamics due to the presence of viscous friction is much faster than the oscillatory dynamics of the spring (i.e., $\tau_{\theta}\ll \tau_{k}$: see~\ref{app:dimensions} for details). In this limit the dynamics of $v$ and $I$ are completely decoupled from that of $x$. Moreover, the reward function~\eqref{eq:power} does not depend on $x$ in any way. We can therefore focus on the variables $v$ and $I$ only, and uniquely characterize the state of the system by the covariances $\moy{v^2}$,  $\moy{vI}$ and  $\moy{I^2}$. The dimension of $\sigma$ reduces to $3$, as well as that of $\lambda$. Equations~\eqref{eq:sigma-dyn} and~\eqref{eq:hamiltonian} can be written in this smaller space by taking (see Eq.~\eqref{eq:matrixesApp} in~\ref{app:dimensions})

\begin{equation}
\label{eq:matrixes}
    M_0 =
    \begin{pmatrix}
        2\beta & 2 & 0 \\
        -1 & \beta+\zeta & 1 \\
        0 & -2 & 2\zeta
    \end{pmatrix}
    \ , \quad
    M_1 = 
    \begin{pmatrix}
        0 & 0 & 0 \\
        0 & 1 & 0 \\
        0 & 0 & 2 \\
    \end{pmatrix}\,,\quad  b=
    \begin{pmatrix}
        2 \\
        0 \\
        0 \\
    \end{pmatrix}\,,\quad  \vectheta=
    \begin{pmatrix}
        0 \\
        0 \\
        1 \\
    \end{pmatrix}
    \ .
\end{equation}

We limit our attention to the case of boundary conditions corresponding to equal stationary states, 
\begin{equation}
\label{eq:bc}
\sigma(0)=\sigma(t_f)=\sigma_s(u_s)\,.
\end{equation}
This case is particularly relevant, because it allows to repeat the process cyclically.
To exploit the perturbative approach outlined in Section~\ref{sec:perturbative}, we also need to choose $u_s$ close enough to $u^*$, and carefully verify \textit{a posteriori} that the protocol $u(t)$ does connect the initial and final values of $\sigma$ also in the original, true dynamics.

When (numerically) solving system~\eqref{eq:system}, one obtains up to 12 different solutions for the unknowns $(u_0,u_f)$. However, most of them are not physically admissible, since they imply complex or negative values of $u$ in the time interval $[0,t_f]$.
In particular, for the range of parameters explored in Fig.~\ref{fig:NetPower}, we only find 2 admissible solutions: we call ``A'' the one closer to $(u_0=0, u_f=0)$ [panel~\ref{fig:NetPower}(a)]\footnote{We actually classify as ``A'' also some solutions, found in the range $1 < u_s/u^* \le 1.02$, which involve negative values of $u_0$ and/or $u_f$, and are therefore not physical.}, ``B'' the farther one [panel~\ref{fig:NetPower}(b)]. Notice that the optimal stationary solution $u=u^*$ is the solution A for the case with boundary conditions $\sigma(0)=\sigma(t_f)=\sigma^*$.

\begin{figure}
\includegraphics[width=.99\linewidth]{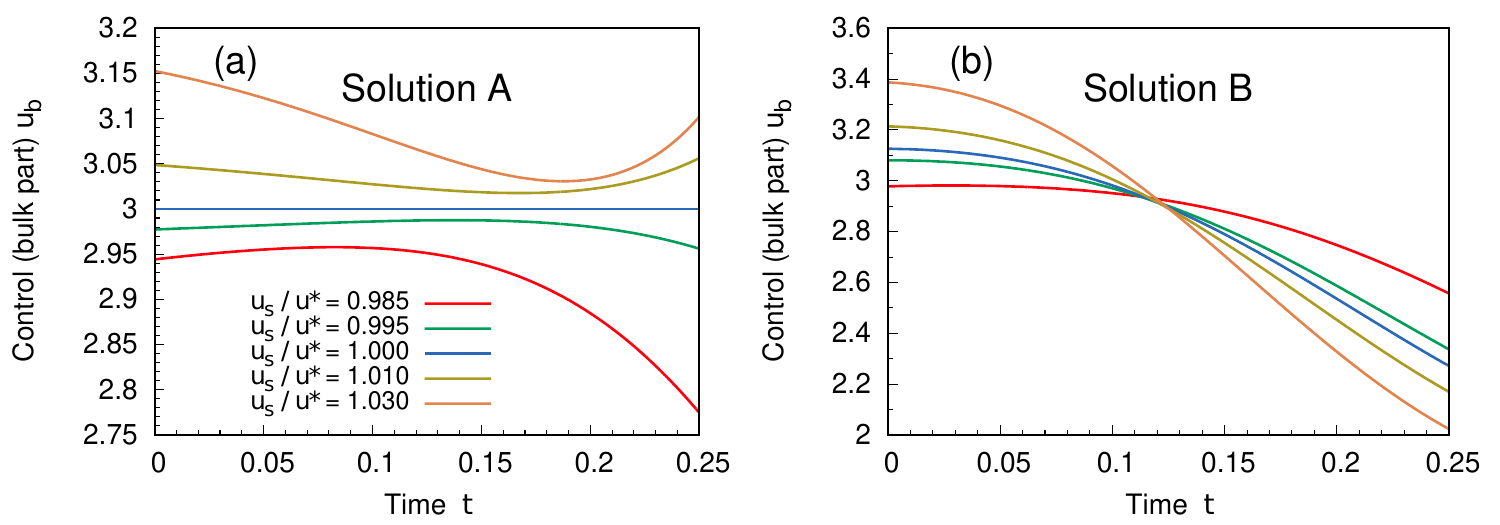}
\caption{Bulk part $u_b$ of the protocol, for the two solutions A [panel (a)] and B [panel (b)], as a function of time. Different boundary conditions are considered. Parameters as in Fig.~\ref{fig:NetPower}.}
\label{fig:Protocols}
\end{figure}

We therefore find that, as shown in Fig.~\ref{fig:NetPower}(c-d), it is possible to find dynamical protocols that perform better than the optimal stationary one. In particular, solution A with $u_s<u^*$ is able to extract larger power than the optimal stationary one, $P^*=P_s(u^*)$. Solution B, in the explored parameter range, is even more performing. We verified that a non-negligible contribution to the extracted power comes from the impulsive changes in $u$ happening at the beginning and at the end of the protocol. Without them, the solutions would not outperform the stationary strategy.

\begin{figure}[ht]
\includegraphics[width=.99\linewidth]{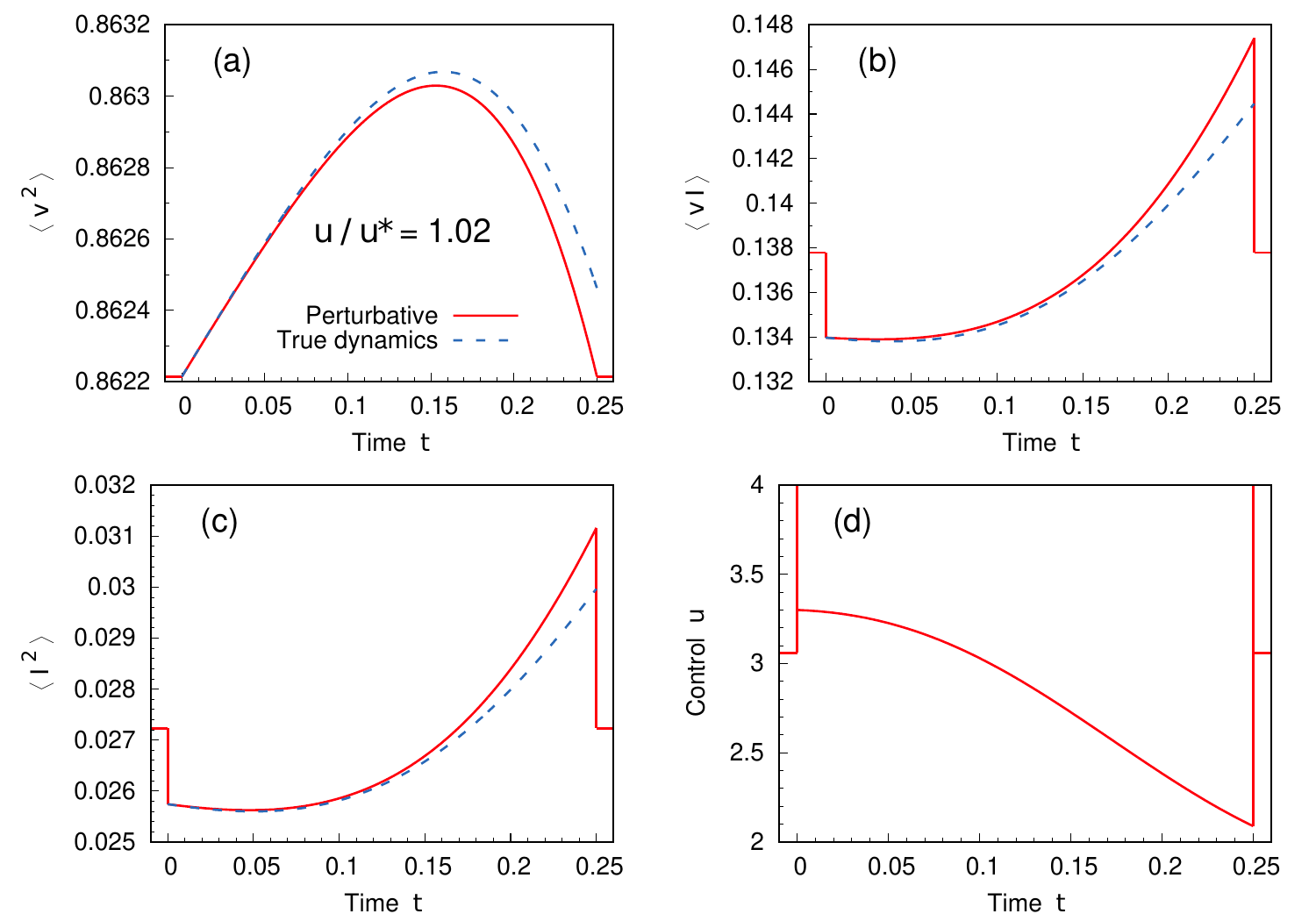}
\caption{Dynamics of the system within solution B, for boundary conditions fixed by $u_s=1.02\, u^*$. Panels (a)-(c) show the evolution of the elements of the vector $\sigma$ (the covariances $\moy{v^2}$, $\moy{v I}$ and $\moy{I^2}$), as computed in the perturbative approach (red solid curves). By inserting the solution protocol $u$, shown in panel (d), back into the original dynamics~\eqref{eq:nodimdyn}, it is possible to compute the true behavior of the system under the prescribed protocol (computation has been done an explicit Runge-Kutta integration scheme of $4-$th order): this evolution is represented by the blue dashed curves in panels (a)-(c). While it is expected that the two sets of curves do not overlap, the fact that they stay close is a consistency check on our perturbative approximation. Parameters as in Fig.~\ref{fig:NetPower}.
\label{fig:dyn}
}
\end{figure}

The bulk part of the protocol, $u_b$, is shown in the two cases in Fig.~\ref{fig:Protocols}. As expected, solution A reduces to the stationary one for $u_s=u^*$. 
The detailed dynamics of solution B for one choice of the boundary conditions is finally shown in Fig.~\ref{fig:dyn}, where the quality of the perturbative approach is also examined.


\section{Discussion}
\label{sec:discussion}
In this work we have presented a proof of concept that dynamical control methods can be used to improve the efficiency of an energy harvester model driven by broadband vibrations. We have considered a linear model, that has been shown to well reproduce the dynamics of electromagnetic and piezoelectric energy harvesters, for which both theoretical and experimental studies are available~\cite{costanzo2021stochastic,Costanzo22entropy}. This kind of systems is described by a Langevin equation for the movable part, coupled with a deterministic equation for the current flowing in the output circuit. We have shown that changing in time the value of the load resistance can lead to larger work extraction.
In particular, we were able to find optimal dynamical solutions by mean of Pontryagin's Principle. In general the prescribed condition leads to a system of ordinary differential equations, quite challenging to solve. However, it is always possible to linearize the equations around the optimal stationary solution, and search for protocols that do not differ too much from it.

The kind of optimal protocols found in this work may appear rather difficult to implement at a practical level, since they involve abrupt discontinuities: we leave for future studies the regularization of these solutions. Such regularization is expected to lead to realistic, implementable protocols, at the price of lower harvesting power. A promising alternative approach that can be pursued on the basis of the present work is to consider optimal updates of the control variable as a function of real-time measurements performed on the system (i.e. closed feedback loop) within the framework of dynamic programming \cite{Bellmann1958}.
\\\\\\
\textbf{Author Contribution}: Conceptualization, A.S. and M.B.; methodology, D.L., A.M., A.P. and M.B.; software, D.L., A.M., A.P. and M.B.; formal analysis, D.L., A.M. and M.B.; investigation, D.L., A.M., A.P., A.S. and M.B.; writing---original draft preparation, D.L., A.M., A.S. and M.B.; writing---review and editing, D.L., A.M., A.P., A.S. and M.B.; visualization, D.L., A.M. and M.B.; supervision, A.S. and M.B. All authors have read and agreed to the published version of the manuscript.
\\\\
\textbf{Data Availability Statement}: Numerical data presented in Section~\ref{sec:results} are available from the authors upon reasonable request.
\\\\
\textbf{Funding}: A.M. acknowledges financial support from the project “MOCA" funded by MUR PRIN2022 grant No. 2022HNW5YL. M.B. was supported by ERC Advanced Grant RG.BIO (Contract No. 785932).
\\\\
\textbf{Conflicts of Interest}: The authors declare no conflicts of interest.
\vspace{6pt} 

\appendix

\section{Energy harvester model in dimensionless units}
\label{app:dimensions}

We discuss here how to rewrite the dynamics~\eqref{eq:harvester} into a more compact form, passing to dimensionless variables.
 We notice that there are 4 dimensional units to be fixed:
\begin{enumerate}
    \item mass $M=1$;
    \item time $\tau= \tau_\theta = \sqrt{M L_c}/\theta=1$;
    \item length $\ell=\sqrt{D_0 \tau^3}=1$;
    \item current intensity $\iota = \theta\ell/L_C=1$.
\end{enumerate}
The dimensionless dynamic variables are then $x^* = x/\ell$, $t^*=t/\tau$, $I^*=I/\iota$. We will drop the asterisks from now on. In these new variables, the dynamics reads
\begin{equation}
    \begin{aligned}
        \dot x &= v \\
        \dot v &= -\alpha x - \beta v - I + \xi \\
        \dot I &= v - \varepsilon I \\
        \moy{\xi(t)\xi(t')} &= 2 \delta(t-t')\,.
    \end{aligned}
\end{equation}
The dimensionless coefficients read
\begin{equation}
    \alpha = \frac{k_s \tau^2_\theta}{M} = \left( \frac{\tau_\theta}{\tau_k} \right)^2 \ , \quad 
    \beta = \frac{\gamma\tau_\theta}{M} = \frac{\tau_\theta}{\tau_v} \ , \quad
    \varepsilon = \frac{(R_C+R)\tau_\theta}{L_C} = \frac{\tau_\theta}{\tau_C} + \frac{\tau_\theta}{\tau_R} = \zeta + u\ ,
\end{equation}
having defined the characteristic times $\tau_k = \sqrt{M/k_s}$ (elastic response), $\tau_v=M/\gamma$ (viscous friction), $\tau_\theta$ (electromechanical coupling), $\tau_C = L_C/R_C$ (coil) and $\tau_R = L_C/R$ (resistance).\\

Just to give an idea of the order of magnitude of these coefficients, we report the experimental values used in~\cite{Costanzo22entropy} in Table~\ref{tab:exp} and the characteristic times and dimensionless parameters in Table~\ref{tab:adim}. We stress that the aim of this paper is not to provide solutions for a realistic system: therefore in our explicit examples we will use instead a more idealized, easier to treat, set of parameters.

\begin{table}[h]
\caption{Experimental values of the electromechanical harvester, measured in ~\cite{Costanzo22entropy}.\label{tab:exp}}
\begin{tabularx}{\textwidth}{cccccc}
    \toprule
    $M$ (\unit{kg}) & $\gamma$ (\unit{kg/s}) & $k_s$ (\unit{N/m}) & $\theta$ (\unit{N/A}) & $L_C$ (\unit{H}) & $R_C$ (\unit{\ohm}) \\
    \num{0.048 +- 0.005} & \num{1.80 +- 0.05} & \num{18810+-50}  & \num{29.9 +- 0.5} & \num{0.124 +- 0.005} & \num{227.6 +- 0.5} \\
    \bottomrule
\end{tabularx}
\end{table}
\begin{table}[h]
\caption{Characteristic times (in seconds) and dimensionless parameters, neglecting errors.\label{tab:adim}}
\begin{tabularx}{\textwidth}{c c c c c c c}
    \toprule
    $\tau_v$ & $\tau_k$ & $\tau_\theta$ & $\tau_C$ & $\alpha$ & $\beta$ & $\zeta$ \\
    \num{2.67e-2} & \num{1.60e-3} & \num{2.58e-3} & \num{5.44e-4} & \num{3.22} & \num{9.66e-2} & \num{4.74} \\
    \bottomrule
\end{tabularx}
\end{table}
Using the Ornstein-Uhlenbeck formalism adopted in~\cite{Costanzo22entropy}, one gets the matrices
\begin{equation}
    \dot X = -A X + \eta \ , \qquad
    A = 
    \begin{pmatrix}
        0 & -1 & 0 \\
        \alpha & \beta & 1 \\ 
        0 & -1 & \varepsilon
    \end{pmatrix} \ , \qquad
    D = 
    \begin{pmatrix}
        0 & 0 & 0 \\
        0 & 1 & 0 \\
        0 & 0 & 0
    \end{pmatrix}
    \ . 
\end{equation}
We stress that in the formalism of Eq.~\eqref{eq:sigma-dyn} and~\eqref{eq:Hamiltonian} this means that the evolution of the state vector
$$
\sigma=\begin{pmatrix}\sigma_{XX} & \sigma_{XV} &\sigma_{XI} & \sigma_{VV} & \sigma_{VI} & \sigma_{II} \end{pmatrix}^T
$$
is determined by 
\begin{equation}
\label{eq:matrixesApp}
    M_0 =
    \begin{pmatrix}
        0 & -2 & 0 & 0 & 0 & 0 \\
        \alpha & \beta & 1 & -1 & 0 & 0 \\
        0 & -1 & \zeta & 0 & -1 & 0 \\
        0 & 2\alpha & 0 & 2\beta & 2 & 0 \\
        0 & 0 & \alpha & -1 & \beta+\zeta & 1 \\
        0 & 0 & 0 & 0 & -2 & 2\zeta
    \end{pmatrix}
    \ , \quad
    M_1 = 
    \begin{pmatrix}
        0 & 0 & 0 & 0 & 0 & 0 \\
        0 & 0 & 0 & 0 & 0 & 0 \\
        0 & 0 & 1 & 0 & 0 & 0 \\
        0 & 0 & 0 & 0 & 0 & 0 \\
        0 & 0 & 0 & 0 & 1 & 0 \\
        0 & 0 & 0 & 0 & 0 & 2 \\
    \end{pmatrix}\,,\quad  b=
    \begin{pmatrix}
        0 \\
        0 \\
        0 \\
        2 \\
        0 \\
        0 \\
    \end{pmatrix}\,,\quad  \vectheta=
    \begin{pmatrix}
        0 \\
        0 \\
        0 \\
        0 \\
        0 \\
        1 \\
    \end{pmatrix}
    \ .
\end{equation}
In the case of constant control $u(t)=u_s$ (and, as a consequence, $\varepsilon(t)=\varepsilon_s$), the stationary covariance matrix $\Sigma_s$ of entries $\sigma^{s}_{\mu \nu} = \moy{X_{\mu} X_{\nu}}$, with $\mu,\nu \in \{X,V,I\}$, satisfies the condition 
$$
2D = A\Sigma_s + \Sigma_s A^T
$$
and the components read
\begin{equation}
    \begin{aligned}
        \sigma^s_{XX} &= \frac{\alpha+\varepsilon(\beta+\varepsilon)}{\alpha[\varepsilon+\beta(1+\alpha+\varepsilon(\beta+\varepsilon))]} \ , & \sigma^s_{XV} &= 0 \ , \\
        \sigma^s_{VV} &= \frac{1+\alpha+\varepsilon(\beta+\varepsilon)}{\varepsilon+\beta(1+\alpha+\varepsilon(\beta+\varepsilon))} \ , & \sigma^s_{XI} &= \frac1{\varepsilon+\beta(1+\alpha+\varepsilon(\beta+\varepsilon))} \ , \\
        \sigma^s_{II} &= \frac1{\varepsilon+\beta(1+\alpha+\varepsilon(\beta+\varepsilon))} \ , & \sigma^s_{VI} &= \frac{\varepsilon}{\varepsilon+\beta(1+\alpha+\varepsilon(\beta+\varepsilon))} \ .
    \end{aligned}
\end{equation}
The stationary extracted power (in dimensionless units) reads
\begin{equation}
    P_s = \frac{\tau^3_\theta}{M\ell^2} R \moy{I^2} = \frac{\tau_\theta}{\tau_R} \moy{I^2} = u \sigma^s_{II} = \frac{u}{\varepsilon+\beta(1+\alpha+\varepsilon(\beta+\varepsilon))}
\end{equation}
and is maximized by the choice (recall that $\varepsilon=\zeta+u$)
\begin{equation}
u^* = \sqrt{\frac{\alpha  \beta +(\beta +\zeta)(1+\beta\zeta) }{\beta }}  \ .
\end{equation}


\section{Boundary conditions and Power Extraction}
\label{app:discontinuity}
Infinite discontinuities in control are convenient for analytical purposes but must be properly interpreted.
Let us consider a control $u(t)$ of the form
\begin{equation}
u(t)=u_0\delta_\epsilon(t-t_0)+u_b(t)+u_f\delta_\epsilon(t-t_f)
\end{equation}
where $0<\epsilon\ll t_f-t_0$. Here we assume that $u_b(t)\neq0$ if and only if $ t\in[t_0+\epsilon,t_f-\epsilon]$. The function $\delta_\epsilon(t)$ denotes a generic distribution fulfilling the property 
\begin{equation}
\int_0^\epsilon {\rm d}t\,\delta_\epsilon(t)=1,
\label{eq:delta_property}
\end{equation}
which converges to the Dirac delta function when $\epsilon\to 0$. As a consequence, in Eq.~\eqref{eq:PMP-sigma} the terms proportional to $u(t)$ represent the leading contribution to $\dot{\sigma}$ for $t\in[t_0,t_0+\epsilon)$ and $t\in(t_f-\epsilon,t_f]$,
that is
\begin{equation}
\dot{\sigma}\simeq -u M_1\sigma\implies \sigma(t)\simeq   \begin{cases}
    e^{-M_1 \int_{t_0}^{t} u(t)}\sigma(t_0)\,;\qquad &t\in[t_0,t_0+\epsilon)\\
 e^{-M_1 \int_{t_f-\epsilon}^{t} u(t)}\sigma(t_f-\epsilon)\,;\qquad &t\in(t_f-\epsilon,t_f]
\end{cases}
\end{equation}
By evaluating the above expressions in $t=t_0+\epsilon$ or $t=t_f$ respectively (using also Eq.\eqref{eq:delta_property}), we obtain 
\begin{align}
    &\sigma(t_0^+)=\lim_{\epsilon\to0} \sigma(t_0+\epsilon) = e^{-M_1 u_0}\sigma(t_0)\\
    &\sigma(t_f^-)=\lim_{\epsilon\to0}\sigma(t_f-\epsilon)= e^{M_1 u_f}\sigma(t_f)\,,
    \label{eq:boundary-linear-appendix}
\end{align}
which coincide with expressions~\eqref{eq:boundary-linear} given in the main text. Let un now focus on the effects of discontinuities on the reward $$J[u] = \int^{t_f}_{t_0} \de t \, u(t) \vectheta\cdot\sigma(t)=
        \vectheta \cdot \left[ \int^{t_f}_{t_0} \de t \, u(t) \sigma(t)\right]\,.$$
Note that the integral above can be split into
\begin{equation}
        \int^{t_f}_{t_0} \de t \, u(t) \sigma(t) = \lim_{\epsilon\to0}
        \left[ \int^{t_0+\epsilon}_{t_0} \de t \, u(t) \sigma(t)+\int^{t_f-\epsilon}_{t_0+\epsilon} \de t \, u(t) \sigma(t)+\int^{t_f}_{t_f-\epsilon} \de t \, u(t) \sigma(t)\right]\,.
        \label{eq:integrals_power}
\end{equation}
Noting that $$u(t)\sigma(t)=-M^{-1}_1\dot{\sigma}$$
for $t\in[t_0,t_0+\epsilon)$ and $t\in(t_f-\epsilon,t_f]$, Eq.~\eqref{eq:integrals_power} can be written as
\begin{align}
        \int^{t_f}_{t_0} \de t \, u(t) \sigma(t) &= \int^{t_f}_{t_0} \de t \, u_b(t) \sigma(t)-\lim_{\epsilon\to0}
        M_1^{-1}\left[ \int^{t_0+\epsilon}_{t_0} \de t \,\dot{ \sigma}(t)+\int^{t_f}_{t_f-\epsilon} \de t \, \dot{\sigma}(t)\right]=\\
        &=\int^{t_f}_{t_0} \de t \, u_b(t) \sigma(t)-M_1^{-1}\left[\sigma(t_0^+)-\sigma(t_0)+\sigma(t_f)-\sigma(t_f^-)\right]\,.
        \label{eq:integral_power_decomposition}
\end{align}
Combining Eqs.~\eqref{eq:boundary-linear-appendix} with Eq.~\eqref{eq:integral_power_decomposition} leads to
\begin{equation}
J[u]=\vectheta \cdot \left[ \int^{t_f}_{t_0} \de t \, u_b(t) \sigma(t)+\left(I-e^{-M_1 u_0}\right)M_1^{-1}\sigma(t_0)+\left(e^{M_1 u_f}-I \right)M_1^{-1}\sigma(t_f)\right]\,,
\end{equation}
which corresponds to Eq.~\eqref{eq:reward_linear} for $\sigma(t_0)=\sigma(t_f)=\sigma_s$.




\bibliographystyle{unsrt}
\bibliography{harvesterbiblio}
\end{document}